\documentclass[twocolumn,showpacs,preprintnumbers,amsmath,amssymb,nofootinbib]{revtex4-1}
\usepackage{bm, color} 
\usepackage{amssymb,amsfonts,slashed,amsthm,amsmath,graphicx, soul}
\usepackage[caption=false]{subfig}
\usepackage{natbib}
\usepackage[utf8]{inputenc}
\newcommand{\GeV}{\  {\rm GeV} }
\newcommand{\TeV}{\  {\rm TeV} }

\newcommand{\lmk}{\left(}  
\newcommand{\rmk}{\right)}

\newcommand{\la}{\left\langle} 
\newcommand{\ra}{\right\rangle}

\newcommand{\Mpl}{M_{\rm Pl}}

\def\REFS#1{Refs.~\cite{#1}}

\begin{document}

\title{
Unification for the Darkly Charged Dark Matter
}

\author{
Ayuki Kamada,$^{1}$
}

\author{
Masaki Yamada,$^{2}$
}

\author{
Tsutomu T. Yanagida$^{3, 4}$
}

\affiliation{
$^{1}$ Center for Theoretical Physics of the Universe, 
Institute for Basic Science (IBS), 55 Expo-ro, Yuseong-gu, Daejeon 34126, Korea
}
\affiliation{
$^{2}$ Institute of Cosmology, Department of Physics and Astronomy, 
Tufts University, 574 Boston Avenue, Medford, MA 02155, U.S.A.
}
\affiliation{
$^{3}$ T. D.  Lee Institute and School of Physics and Astronomy, Shanghai Jiao Tong University, 800 Dongchuan Rd, Shanghai 200240, China
}
\affiliation{$^{4}$ Kavli IPMU (WPI), UTIAS, 
The University of Tokyo, 5-1-5 Kashiwanoha, Kashiwa, Chiba 277-8583, Japan
}

\begin{abstract}
We provide a simple UV theory for a Dirac dark matter with a massless Abelian gauge boson. We introduce a single fermion transforming as the $\bm{16}$ representation in the SO(10)$'$ gauge group, which is assumed to be spontaneously broken to SU(5)$'\times$U(1)$'$. The SU(5)$'$ gauge interaction becomes strong at an intermediate scale and then we obtain a light composite Dirac fermion with U(1)$'$ gauge interaction at the low-energy scale. Its thermal relic can explain the observed amount of dark matter consistently with other cosmological and astrophysical constraints. We discuss that a nonzero kinetic mixing between the U(1)$'$ gauge boson and the Hypercharge gauge boson is allowed and the temperature of the visible sector and the dark matter sector can be equal to each other. 
\end{abstract}

\maketitle

{\bf Introduction.--}
Constructing a grand unified theory (GUT) of the Standard Model (SM) is an outstanding challenge in particle physics. 
The similarity of the SM gauge coupling constants and the beautiful unification of fermions in the SU(5) multiplets may support the existence of the unified theory at a very high energy scale. 
However, the running of the gauge coupling constants 
and the quark/lepton mass relation are deviated from the simplest SU(5) GUT prediction~\cite{Ellis:1990zq,Ellis:1990wk,Amaldi:1991cn,Giunti:1991ta,Langacker:1991an}, 
which may imply that the GUT breaking in the visible sector is much more complicated than we expect.

In the context of cosmology, there exists dark matter, which may be a fundamental particle 
that barely interacts with the SM particles. Since the dark matter (DM) must be stable and neutral under the electromagnetic interaction, we consider it to be charged under a hidden U(1)$'$ gauge symmetry. 
Then one may hope that the dark sector is also unified into a GUT$'$ theory as in the SM sector. 

In this letter, we propose a chiral SO(10)$\times$SO(10)$'$ GUT as a unified model of SM and DM sectors. 
The first SO(10) gauge theory is a standard SO(10) GUT model, which we do not specify as it has been extensively discussed in the literature~\cite{Chang:1984qr,Chang:1984qr,Deshpande:1992au,Fukugita:2003en,Bertolini:2009es,Mambrini:2013iaa,Mambrini:2015vna}. 
We focus on the second SO(10)$'$ gauge theory, which gives a dark sector. 
The fermionic matter content in SO(10)$'$ is a single field in the $\bm{16}$ representation. 
The SO(10)$'$ is assumed to be spontaneously broken to SU(5)$'$$\times$U(1)$'$ at a very high energy scale 
and the SU(5)$'$ gauge interaction becomes strong at the energy scale of order $10^{13} \GeV$. 
Below the confinement scale, we have a light composite Dirac fermion charged under the remaining U(1)$'$. 
Therefore the DM sector results in a Dirac DM with a massless U(1)$'$ gauge boson, which has been discussed 
in Refs.~\cite{Feng:2009mn, Agrawal:2016quu}. 
A similar idea of the strong SU(5)$'$ gauge theory 
was used in the literature in different contexts~\cite{ArkaniHamed:1998pf, Gavela:2018paw, Kamada:2019gpp}, 
where they did or did not introduce the U(1)$'$ gauge symmetry.

As discussed in Ref.~\cite{Agrawal:2016quu}, 
a DM with a massless hidden photon is still allowed by any astrophysical observations and DM constraints 
even if it is the dominant component of DM.  
The thermal relic abundance of the Dirac fermion can explain the observed amount of DM. 
We find that 
the temperatures of SM and DM sectors can be the same with each other at a high temperature. 
This allows us to consider a 
nonzero kinetic mixing between the U(1)$'$ and U(1)$_Y$ gauge bosons, 
which presents an interesting possibility for the DM search in this model. 
The relic of the massless U(1)$'$ gauge boson affects the expansion rate of the Universe as dark radiation, 
which can be checked by the detailed measurements of the CMB anisotropies in the future.

\vspace{0.3cm}
{\bf Dark matter in the low-energy sector.--}
We first explain a low energy phenomenology in the dark sector. 
Let us introduce a U(1)$'$ gauge symmetry and 
a Dirac fermion $\eta$ of weak-scale mass $m_\eta$ with charge $q$. 
We consider the case where the U(1)$'$ gauge symmetry is not spontaneously broken 
and the gauge boson $\gamma'$ is massless until present. 
We denote the temperature of dark sector as ${T'}$ 
and that of visible sector as $T$. 
We define $\xi (T)= T' / T$, which depends on the temperature. 
We will see that there is a viable parameter region even if $\xi = 1$ at a high temperature.

\begin{figure}[t]
  \begin{center}
    \includegraphics[keepaspectratio=true,width=85mm]{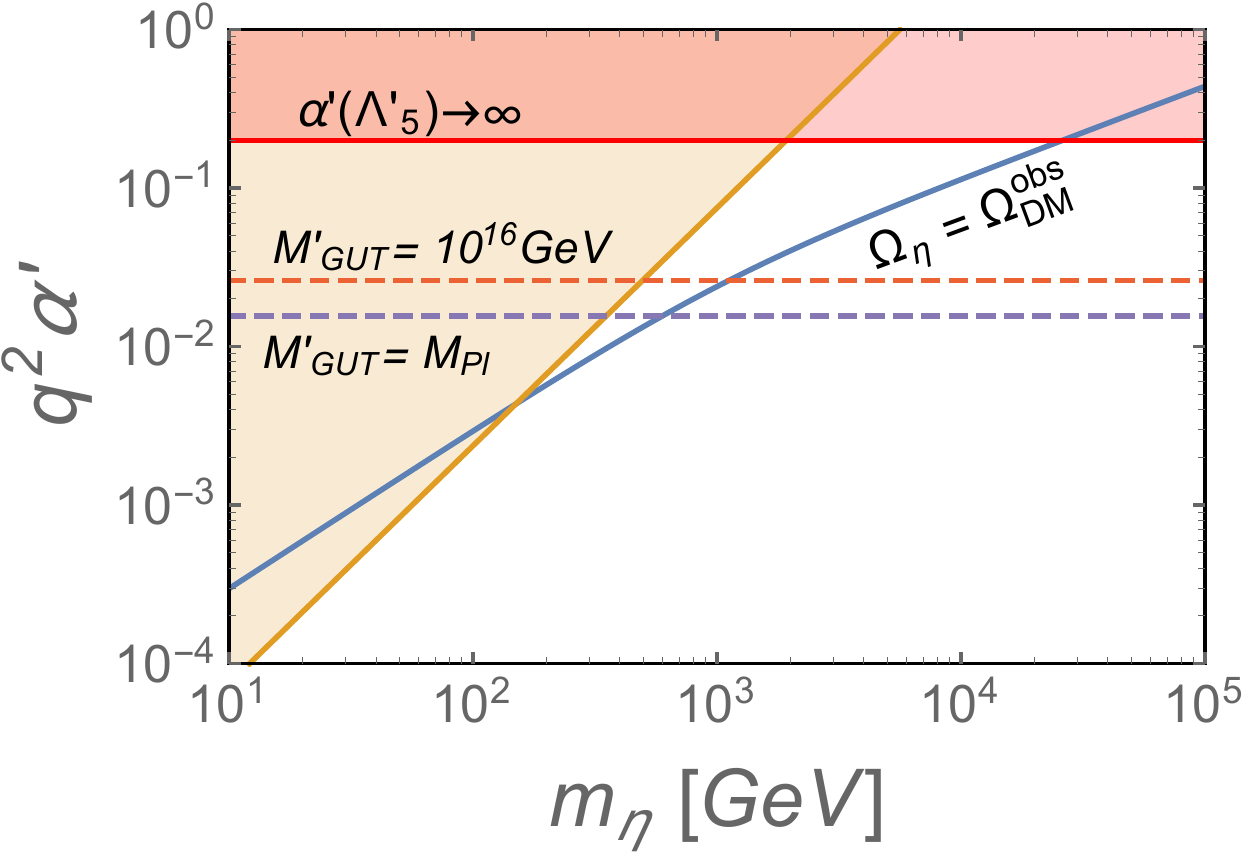}
  \end{center}
  \caption{
Constraint on $q^2 {\alpha'}$ as a function of $m_\eta$. 
We can explain the observed amount of DM on the solid blue curve when $\xi = 1$ at the time of DM freeze out. 
The orange shaded region is excluded by the ellipticity constraint on the observed galaxy. 
On the upper and lower dashed lines, 
the gauge coupling constant can be unified with that of the SU(5)$'$ gauge symmetry at the energy scales of $M_{\rm GUT}' =10^{16} \GeV$ and the Planck scale, respectively, for the case of $q = \sqrt{10} / 4$. 
Above the red line, the gauge coupling constant blows up at the energy scale below $\Lambda'_5 = 10^{13} \GeV$. 
}%
  \label{fig1}
\end{figure}

The DM can annihilate into the dark photon 
and hence its thermal relic density is determined by the freeze-out process. 
The thermally-averaged annihilation cross section is given by 
\begin{eqnarray}
 \la \sigma v_{\rm Mol} \ra = \frac{\pi q^4 {\alpha'}^2}{m_\eta^2} \bar{S}_{\rm ann} (\alpha'), 
\end{eqnarray}
where $v_{\rm Mol}$ is Moller velocity 
and 
$\bar{S}_{\rm ann}$ is the thermally-averaged Sommerfeld enhancement factor~\cite{Gondolo:1990dk, vonHarling:2014kha}. 
In the regime where the gauge interaction is relatively large, 
a bound-state formation is efficient and is relevant to determine the thermal relic abundance. 
Hence we have to solve the coupled Boltzmann equations for the unbound and bound DM particles 
as done in Ref.~\cite{vonHarling:2014kha}. 
In Fig.~\ref{fig1}, 
we quote their result to plot a contour on which we can explain the observed amount of DM 
for the case of $\xi (T)= 1$ at the time of DM freeze-out.

The DM has a self-interaction mediated by the dark photon. 
Its cross section is given by 
\begin{eqnarray}
 \frac{\sigma_T}{m_\eta} 
 &=& \frac{8 \pi {\alpha'}^2}{m_\eta^3 v^4} \log \Lambda 
 \\
 &\simeq& 
 0.2 \ {\rm cm}^2 / {\rm g} 
 \lmk \frac{q^2 \alpha'}{0.025} \rmk^2 
 \lmk \frac{m_\eta}{1 \TeV} \rmk^{-3} 
 \lmk \frac{v}{300 \ {\rm km/s} } \rmk^{-4}, 
 \nonumber\\
\end{eqnarray}
where $\log \Lambda$ ($\approx 40$ - $70$) comes from an infrared cutoff for the scattering process. 
The velocity of DM $v$ depends on the scale we are interested in: 
$v \sim 30 \ {\rm km/s}$, $300 \ {\rm km/s}$, and $1000 \ {\rm km/s}$ for dwarf galaxies, galaxies, and galactic clusters, respectively. 
The observed triaxial structure of a galaxy NGC720 puts a stringent upper bound on the self-interaction cross section since the DM velocity distribution is randomized and is more isotropic by the self-interaction~\cite{Buote:2002wd, Feng:2009mn, Agrawal:2016quu}. 
This can be rewritten as a constraint on the gauge coupling constant 
and is shown as the orange shaded region in Fig.~\ref{fig1}. 
The DM with mass of order $0.1$ - $10 \TeV$ 
is allowed even if $\xi = 1$ at the time of freeze-out, 
depending on $q^2 {\alpha'}$ ($\gtrsim 10^{-2}$). 
We expect that a larger number of statistical samples of galactic structures 
will make the analysis more robust.

Since the self-interacting cross section is proportional to $v^{-4}$, 
the cross section for the cluster scales is much smaller than the observational constraints~\cite{Kaplinghat:2015aga}. 
On the other hand, 
the self-interaction is quite large in the smaller scales, like dwarf galaxies. 
It has been discussed that 
a too large scattering cross section leads to a very short mean-free path, 
which suppresses heat conduction and hence both core formation and core collapse are inhibited~\cite{Ahn:2002vx, Ahn:2004xt}. 
Therefore, the constraint on the dwarf galactic scales may not be applied to this kind of models 
and 
the massless mediator is still allowed for the self-interacting DM model.

The massless dark photon remains in the thermal plasma in the dark sector 
and contributes to the energy density of the Universe 
as dark radiation. 
Its abundance is conveniently described by the deviation of the effective neutrino number 
from the SM prediction such as 
\begin{eqnarray}
 \Delta N_{\rm eff} 
 &=& \frac{8}{7} \lmk \frac{2}{g'_{*} ({T'_{\rm d}})} \frac{g_*({T'_{\rm d}})}{43/4}  \rmk^{-4/3} \xi^4 ({T'_{\rm d}}), 
 \label{Neff}
\end{eqnarray}
where $g'_{*}$ is the effective number of degrees of freedom in the dark sector 
and ${T'_{\rm d}}$ is the decoupling temperature of dark sector from the SM sector. 
In the case where the dark sector is completely decoupled from the SM sector before the DM becomes non-relativistic and the electroweak phase transition, 
we should take $g'_{*} ({T'_{\rm d}}) = 2+ 4 (7/8) = 11/2$ and $g_*({T'_{\rm d}}) = 106.75$ 
and obtain $\Delta N_{\rm eff} = 0.21 \xi^4 ({T'_{\rm d}})$. 
Even if we set $\xi ({T'_{\rm d}})= 1$, 
the prediction is consistent with the constraint reported by 
the Planck data combined with the BAO observation: 
$N_{\rm eff} = 3.27 \pm 0.15$~\cite{Aghanim:2018eyx}. 
We can check the deviation from the SM prediction 
with a large significance in the near future by, e.g., the CMB-S4 experiment~\cite{Wu:2014hta, Abazajian:2016yjj}.

It is also possible that 
the DM sector is in the thermal equilibrium with the SM sector at a high temperature 
and then decoupled after the DM becomes non-relativistic. 
This is the case when the U(1)$'$ gauge boson has a nonzero kinetic mixing with the U(1)$_Y$ gauge boson as we will discuss later. 
Then we should take $\xi ({T'_{\rm d}}) = 1$ and 
$g'_{*} ({T'_{\rm d}}) = 2$. 
As we will discuss shortly, 
the decoupling temperature is just below the DM mass, which is of order or larger than the electroweak scale. 
Thus we expect 
$g_{*} ({T'_{\rm d}}) \simeq 100$, 
which results in $\Delta N_{\rm eff} \simeq 0.07$. 
This scenario is also consistent with the Planck data 
and would be checked by the CMB-S4 experiment in the future.

\vspace{0.3cm}
{\bf Dark matter from hidden $\bm{SO(10)}'$.--}
Now we shall provide a UV theory of the DM sector, which is similar to the SM GUT. 
We introduce an SO(10)$'$ gauge group and a chiral fermion transforming as the $\bm{16}$ representation, 
assuming that the gauge group is spontaneously broken to SU(5)$'$$\times$U(1)$'$ 
at the energy scale much above $10^{13} \GeV$ and below the Planck scale. 
After the SSB, the fermion is decomposed into 
$\psi$, $\chi$, and $N$, which transform as the
$\bm{\bar{5}}$, $\bm{10}$, and $\bm{1}$ representations in the SU(5)$'$ gauge group, respectively. 
If we denote the U(1)$'$ charge of $N$ as $q$ ($= \sqrt{10}/4$), 
those of $\psi$ and $\chi$ are $-3 q / 5$ and $q/5$, respectively~\cite{Pacholek:2013kca}. 
If one starts from a generic SU(5)$'$$\times$U(1)$'$ gauge theory instead of the SO(10)$'$ gauge theory, 
the U(1)$'$ charge $q$ may be different from $\sqrt{10}/4$.

Since the SU(5)$'$ gauge interaction is asymptotically free, 
it becomes strong and is confined at a dynamical scale $\Lambda'_5$. 
Below the confinement scale, 
there is a massless baryonic state composed of three fermions like $\eta = \psi \psi \chi$ 
as the t'Hooft anomaly matching condition is satisfied~\cite{tHooft:1979rat, Dimopoulos:1980hn} (see Refs.~\cite{ArkaniHamed:1998pf, Gavela:2018paw, Kamada:2019gpp} for other applications of this model). 
This can be combined with $N$ to form a Dirac fermion. 
In fact, we can write down the following dimension-6 operator: 
\begin{eqnarray}
 \frac{c}{\Mpl^2} \psi \psi \chi N + {\rm h.c.}, 
\end{eqnarray}
where $c$ is an ${\cal O}(1)$ constant. 
This results in a Dirac mass term below the dynamical scale 
and its mass is roughly given by 
\begin{eqnarray}
 m_\eta \sim c \frac{(\Lambda'_5)^3}{\Mpl^2}. 
\end{eqnarray}
This is of order $100 \GeV - 10 \TeV$ when the dynamical scale $\Lambda'_5$ is of order $10^{13 \, \text{-}\, 14} \GeV$. 
As a result, the low-energy sector is nothing but the DM model discussed in the previous section.

As for the SM sector, we consider also an SO(10) GUT, 
motivated by the thermal leptogenesis~\cite{Fukugita:1986hr} (see, e.g., \REFS{Buchmuller:2002rq, Giudice:2003jh, Buchmuller:2005eh, Davidson:2008bu} for recent reviews) 
and seesaw mechanism~\cite{Minkowski:1977sc, Yanagida:1979as, GellMann:1980vs, Glashow:1979nm}. 
Here, we introduce a right-handed neutrino with mass of order or larger than $10^9 \GeV$ in the SM sector. 
Then, we expect an SO(10)$\times$SO(10)$'$ gauge theory to be a unified model of the SM and DM sectors. 
The similarity of the SM and DM sectors may be because a fermion in the $\bm{16}$ representation is the minimal particle content for the anomaly-free chiral SO(10) gauge theory.

\begin{figure}[t]
  \begin{center}
    \includegraphics[keepaspectratio=true,width=85mm]{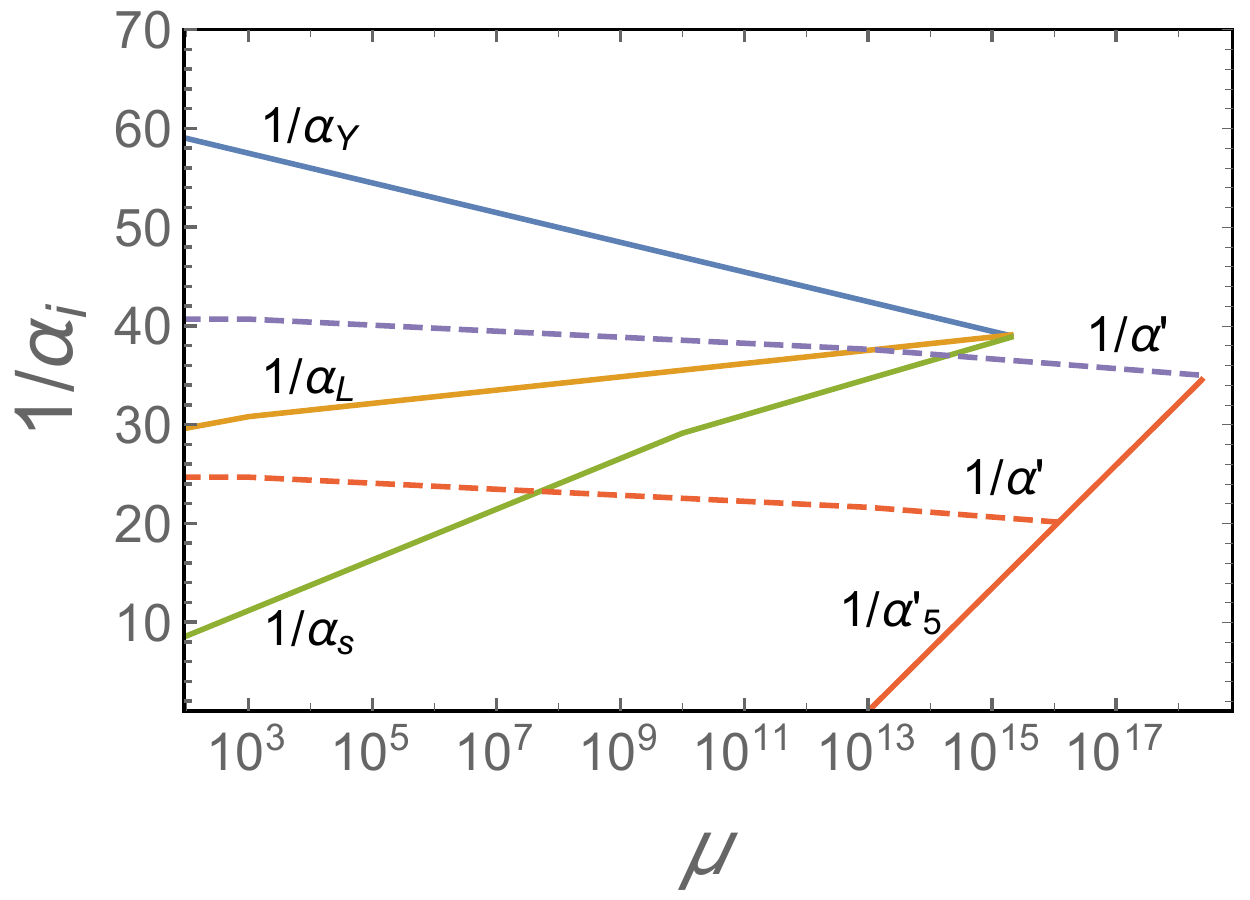}
  \end{center}
  \caption{
Renormalization group running of gauge coupling constants, 
where $\mu$ is the renormalization scale in units of GeV. 
We introduce adjoints fermions for SU(2)$_L$ and SU(3)$_c$ at $10^{3}\GeV$ and $10^{10}\GeV$, respectively, to present an example of gauge coupling unification of the SM gauge interactions~\cite{Aizawa:2014iea}. 
We plot the running of ${\alpha'}$ with $q = \sqrt{10}/4$ for the case in which it is unified with SU(5)$'$ gauge coupling constant $\alpha'_5$ at the energy scale of $10^{16} \GeV$ (red dashed line) and the Planck scale (blue dashed line). 
}%
  \label{fig2}
\end{figure}

An example of renormalization group running of gauge coupling constants is shown in Fig.~\ref{fig2}, 
where we note that there are three flavors for quarks and leptons while there is only one ``flavor" in the dark sector. 
Although an explicit construction of the GUT model in the SM sector is beyond the scope of this paper, 
we present a gauge coupling unification in a simple GUT model proposed in \cite{Aizawa:2014iea}. 
They introduced 
adjoint fermions for SU(3)$_c$ and SU(2)$_L$ at an intermediate scale and at the TeV scale, respectively. 
Although the SU(2)$_L$ adjoint fermion is stable, 
we assume that it is a subdominant component of DM or 
there is another field that makes it unstable. 
Noting that this is just one example of GUT in the Standard Model sector, 
we plot the gauge coupling unification in the simplest case in the figure. 
We do not introduce such adjoint fermions in the dark sector 
or we assume that they are heavier than the dynamical scale if present.

We are interested in the case where 
$q=\sqrt{10}/4$ and 
the SU(5)$'$ gauge coupling $\alpha'_5$ becomes strong at $\Lambda'_5 \sim 10^{13} \GeV$. 
Starting from ${\alpha'} \simeq 4.2 \times 10^{-2}$ and $2.5 \times 10^{-2}$ at the electroweak scale, 
we find that the SU(5)$'$$\times$U(1)$'$ gauge group can be unified at the energy scale of $M_{\rm GUT}' = 10^{16} \GeV$ and the 
Planck scale, respectively. 
These gauge coupling constants are shown as the upper and lower dashed lines in Fig.~\ref{fig1}. 
It shows that the DM mass should be about $1.1 \TeV$ and $600 \GeV$, respectively, 
to explain the observed amount of DM if $\xi (T_{\rm d}') = 1$.

We note that the gauge coupling constants in the dark sector does not need to be unified at the same scale as the GUT scale in the SM but can be unified at the energy scale 
between the dynamical scale $\Lambda'_5$ ($\sim 10^{13} \GeV$) 
and the Planck scale. 
Thus the U(1)$'$ gauge coupling constant can be as large as $q^2 {\alpha'} \sim 0.2$ at the electroweak scale. 
However, we expect that 
the gauge coupling constant at the unification scale is 
of the same order with that of the SM gauge coupling constants 
and hence $M_{\rm GUT}' = {\cal O}(10^{16 \, \text{-} \, 18}) \GeV$. 
In this case, $\alpha'$ must be within the region between the dashed lines in Fig.~\ref{fig1}, 
namely, 
\begin{eqnarray}
 \alpha' = (2.5 \, \text{-} \, 4.2) \times 10^{-2}, 
 \quad m_\eta = 0.6 \, \text{-} \, 1.1 \TeV. 
\end{eqnarray}
This is the prediction of the chiral SO(10)$'$ gauge theory in the DM sector.

\vspace{0.3cm}
{\bf Kinetic mixing.--}
Finally, we comment on the kinetic mixing between the U(1)$_Y$ and U(1)$'$ gauge bosons. 
For this purpose, we need to specify how to break the gauge groups at the GUT scale. 
We first note that a scalar field transforming as the $\bm{45}$ representation in SO(10) is decomposed into scalar fields in the $\bm{1} + \bm{10} + \bar{\bm{10}} + \bm{24}$ representations under an SU(5) $(\subset SO(10))$ gauge group. The singlet $\bm{1}$ can be used to break SO(10) to SU(5)$\times$U(1). 
We assume that 
SO(10) and SO(10)$'$ are spontaneously broken to 
SU(5)$\times$ U(1)$_{(B-L)}$ and SU(5)$'$ $\times$U(1)$'$ 
by nonzero VEVs of 
$\bm{45}_H$ and $\bm{45}'_H$, respectively. 
The remaining SU(5) in the visible sector is also assumed to be spontaneously broken to the Standard Model gauge group G$_{\rm SM}$ by the field in the $\bm{24}$ representation that is contained in $\bm{45}_H$. 
On the other hand, we assume that $\bm{24}'$ in $\bm{45}'_H$ has a vanishing VEV. 
We finally obtain G$_{\rm SM} \times $U(1)$_{(B-L)}\times$SU(5)$'\times$U(1)$'$ below these energy scales. 
The U(1)$_{(B-L)}$ is assumed to be spontaneously broken at an intermediate scale to give a nonzero mass to the right-handed neutrinos. 

Then 
even if we start from the SO(10)$\times$SO(10)$'$ gauge theory, 
the kinetic mixing between U(1)$_Y$ and U(1)$'$ is induced from the following dimension 6 operator: 
\begin{eqnarray}
 \frac{c'}{\Mpl^2} \bm{45}_H \lmk F_{10} \rmk_{\mu \nu}  \bm{45}'_H \lmk F'_{10'} \rmk^{\mu \nu}
\end{eqnarray}
where $c'$ is an ${\cal O}(1)$ constant, 
$F_{10}$ and $F'_{10'}$ are field strengths of SO(10) and SO(10)$'$, respectively. 
The kinetic mixing parameter is of order $\epsilon \sim c' (v_{\rm GUT} / \Mpl) (v' / \Mpl)$, 
where $v_{\rm GUT}$ and $v'$ are the VEVs of $\bm{24}$ ($\subset \bm{45}_H$) and $\bm{45}'_H$, respectively. 
We expect that the hidden SO(10)$'$ is spontaneously broken 
between the energy scale of $10^{16} \GeV$ and the Planck scale. 
Therefore the factor of $v'/\Mpl$ can be ${\cal O}(10^{-2}$ - $1)$ 
and hence $\epsilon$ is ${\cal O}(10^{-(3 \, {\text - }\, 6)})$ 
for $c' = 0.1$ - $1$.

The dark photon $\gamma'$ can be in thermal equilibrium 
with the SM sector 
by the annihilation and inverse-annihilation processes of DM into the SM particles 
$f \bar{f} \leftrightarrow \eta \bar{\eta}$, the Compton scattering process $\eta \gamma \leftrightarrow \eta \gamma (\gamma')$, 
and the Coulomb scattering process $f \eta \leftrightarrow f \eta$ 
via the kinetic mixing, 
where $f$ represents generic SM particles with nonzero U(1)$_Y$ charges. 
Comparing the energy transfer rate $\Gamma$ 
with the Hubble expansion rate $H$, 
we find that 
the these processes are most important at the temperature around the DM mass. 
The ratio at $T \sim m_\eta$ is roughly given by 
\begin{eqnarray}
 \frac{\Gamma}{H} 
 \sim 
 \frac{\epsilon^2 q^2 \alpha {\alpha'} n_f}{m_\eta^2 H(m_\eta)} 
 \sim \lmk \frac{\epsilon}{10^{-6}} \rmk^2 
 \lmk \frac{q^2 \alpha'}{0.02} \rmk
  \lmk \frac{m_\eta}{1 \TeV} \rmk^{-1},~~ 
 \label{thermaleq}
\end{eqnarray}
where $n_f$ is the number density of the SM particles with nonzero U(1)$_Y$ charges. 
The ratio is larger than of order unity when $\epsilon \gtrsim 10^{-6}$ 
for $m_\eta = 1 \TeV$. 
This process freezes out soon after the DM becomes non-relativistic, 
that is, 
around the temperature of order ${\cal O}(0.1) m_\eta$. 
Therefore, if the kinetic mixing is not strongly suppressed, 
the temperature of the DM sector is the same as the SM sector around the time of DM freeze-out 
and we should take $\xi ({T'_{\rm d}}) = 1$.

The nonzero kinetic mixing between 
the U(1)$_Y$ (or U(1)$_{\rm EM}$) 
and U(1)$'$ gauge bosons leads to a rich phenomenology 
for the DM detection experiments. 
It is convenient to diagonalize the gauge bosons 
in the basis that the SM particles are charged only under U(1)$_{\rm EM}$ 
and the DM is charged under both U(1)$_{\rm EM}$ and U(1)$'$. 
The effective electromagnetic charge of DM is given by 
$q_{\rm eff} = - \epsilon q e' {\rm cos} \theta_W / e_{\rm EM}$, where $e_{\rm EM}$ 
is the gauge coupling of U(1)$_{\rm EM}$ 
and $\theta_W$ is the Weinberg angle. 
The direct detection experiments for DM put a stringent constraint on such a millicharged DM~\cite{Akerib:2013tjd, Aprile:2018dbl}. 
However, 
the constraint is not applicable to the DM with a relatively large charge 
because the DM loses its kinetic energy in the atmosphere~\cite{Dimopoulos:1989hk}. 
The measurement of CMB temperature anisotropies also constrain the millicharged DM 
for a larger charge region~\cite{Kamada:2016qjo, Xu:2018efh}. 
In combination, there is an allowed range such as%
\footnote{
A much stronger constraint may be derived by requiring that the millicharged DM does not diffuse within galactic clusters~\cite{Kadota:2016tqq}, though simulations may be required to correctly take into account the nonlinear gravity effect~\cite{Sepp:2016tfs, Alvis:2018yte}. }
\begin{eqnarray}
 10^{-6}\lmk \frac{m_\eta}{10^{3} \GeV} \rmk \lesssim \epsilon \lesssim 3 \times10^{-5} \lmk \frac{m_\eta}{10^{3} \GeV} \rmk^{1/2}. 
\end{eqnarray}
This can be consistent with the SO(10)$\times$SO(10)$'$ model because 
$\epsilon = {\cal O}(10^{-(3 \, {\text -}\, 6)})$ depending on the SSB scale of SO(10)$'$.

Finally, we comment on the case in which the kinetic mixing is as small as $10^{-(10 \, {\text -} \, 11)}$. Such a small kinetic mixing can be realized if there is Pati-Salam symmetry for the SM sector at an intermediate scale 
and the VEV of $\bm{24}$ $(\subset \bm{45}_H)$ is much smaller than the GUT scale, or $c' \simeq 10^{-6}$. 
In this case, the DM sector is completely decoupled from the SM sector 
even in the early Universe 
and the ratio of the temperatures in these sectors is determined solely by the branching ratio of the inflaton decay into these sectors. 
We note that the gauge-coupling--mass relation of DM, which is shown as the blue curve in Fig.~\ref{fig1}, changes only of order $\sqrt{\xi (T_d')}$ 
unless the Sommerfeld enhancement effect is strongly efficient. 
The constraint by the direct detection experiment of DM for such a very small kinetic mixing is 
given by $\epsilon \lesssim 10^{-10} ( m_\eta / 1\TeV)^{1/2}$ for $m_\eta \gtrsim 100 \GeV$~\cite{Aprile:2018dbl, Hambye:2018dpi}. 
This constraint will be improved by LZ experiment for 1000 days by a factor of about $10$~\cite{Akerib:2018lyp}.

\vspace{0.3cm}
{\bf Discussion.--}
We have proposed a chiral SO(10)$'$ gauge theory as a UV theory of a light Dirac DM that is charged under the hidden U(1)$'$ gauge symmetry. 
A darkly-charged DM is also considered as the double-disk-DM, though it must be a subdominant component~\cite{Fan:2013yva, Fan:2013tia, McCullough:2013jma}. 
A similar model with a nonzero kinetic mixing between U(1)$'$ and the electroweak U(1) gauge bosons, namely the millicharged (or mini-charged) DM model, 
is also motivated by the absorption profile around $78 \, {\rm MHz}$ in the sky-averaged spectrum of 21 cm line by EDGES experiment~\cite{Bowman:2018yin, Munoz:2018pzp, Berlin:2018sjs, Barkana:2018cct, Slatyer:2018aqg, Liu:2018uzy, Kovetz:2018zan}. 
The DM with a massive U(1)$'$ gauge boson is also considered in Refs.~\cite{Tulin:2012wi, Dasgupta:2013zpn, Bringmann:2013vra, Ko:2014bka, Cherry:2014xra, 
Kaplinghat:2015aga, 
Kitahara:2016zyb, Ma:2017ucp, Balducci:2018ryj, Kamada:2018zxi, Kamada:2018kmi, Kamada:2019gpp}. 
Our SO(10)$'$ gauge theory may also be a natural candidate for the UV theory of those models.

The DM has a self-interaction mediated by the gauge boson. 
The cross section is velocity dependent, which is supported by the observations of DM halos in galaxy and galaxy cluster scales. 
As the DM couples to the SM sector only via the small kinetic mixing, 
the gravitational search is one of the important DM searches in our model (see, e.g., Ref.~\cite{Buckley:2017ijx}). 
It would be interesting to collect a larger number of samples 
in different length scales so that 
we can determine the velocity dependence on the self-interaction cross section~\cite{Kaplinghat:2015aga, Tulin:2017ara}. 
This may allow us to distinguish our model from the self-interacting DM model with a velocity-independent cross section, like the ones studied in Refs.~\cite{Hochberg:2014dra, Hochberg:2014kqa, Lee:2015gsa, Hochberg:2015vrg, Kamada:2016ois, Choi:2017zww}. 
It is also worth to investigate if the self-interacting DM with a massless vector mediator solves the small-scale issues for the cosmological structure formation~\cite{Balberg:2002ue, Ahn:2004xt, Koda:2011yb, Essig:2018pzq}.

%
\vspace{0.2cm}
{\bf Acknowledgments.--}
%
A.~K. was supported by Institute for Basic Science under the project code, IBS-R018-D1.
A.~K. would like to acknowledge the Mainz Institute for Theoretical Physics (MITP) of the Cluster of Excellence PRISMA+ (Project ID 39083149) for enabling A.~K. to complete a significant portion of this work.
T.~T.~Y. was supported in part by the China Grant for Talent Scientific Start-Up Project and the JSPS Grant-in-Aid for Scientific Research No.~16H02176, No.~17H02878, and No.~19H05810 and by World Premier International Research Center Initiative (WPI Initiative), MEXT, Japan.
T.~T.~Y. thanks to Hamamatsu Photonics.

\bibliography{reference}

\end{document}